\documentclass[conference]{IEEEtran}
\DeclareUnicodeCharacter{2074}{$^4$}
\IEEEoverridecommandlockouts
\usepackage{cite}
\usepackage{amsmath,amssymb,amsfonts}
\usepackage{algorithmic}
\usepackage{graphicx}
\usepackage{textcomp}
\usepackage{xcolor}
\def\BibTeX{{\rm B\kern-.05em{\sc i\kern-.025em b}\kern-.08em
    T\kern-.1667em\lower.7ex\hbox{E}\kern-.125emX}}
\begin{document}

\title{Performance Review on LLM for solving leetcode problems\\
\thanks{Identify applicable funding agency here. If none, delete this.}
}

\author{\IEEEauthorblockN{1\textsuperscript{st} Lun Wang*}
\IEEEauthorblockA{\textit{Duke University} \\
North Carolina, USA \\
lun.wang@alumni.duke.edu}
\and
\IEEEauthorblockN{2\textsuperscript{nd} Chuanqi Shi}
\IEEEauthorblockA{\textit{University of California San Diego} \\
California, USA \\
chs028@ucsd.edu.com}
\and
\IEEEauthorblockN{3\textsuperscript{rd} Shaoshuai Du}
\IEEEauthorblockA{\textit{University of Amsterdam} \\
Amsterdam, Netherlands \\
s.du@uva.nl}
\and
\IEEEauthorblockN{4\textsuperscript{th} Yiyi Tao}
\IEEEauthorblockA{\textit{Johns Hopkins University} \\
Maryland, USC \\
ytao23@jhu.edu}
\and
\IEEEauthorblockN{5\textsuperscript{th} Yixian shen}
\IEEEauthorblockA{\textit{University of Amsterdam} \\
Amsterdam, Netherlands \\
y.shen@uva.nl}
\and
\IEEEauthorblockN{6\textsuperscript{th} Hang Zhang}
\IEEEauthorblockA{\textit{University of California San Diego} \\
California, USA \\
haz006@ucsd.edu}
\and
\IEEEauthorblockN{7\textsuperscript{th} Yanxin Shen}
\IEEEauthorblockA{\textit{Simon Fraser University} \\
Burnaby, Canada \\
yanxin\_shen@sfu.ca}
\and
\IEEEauthorblockN{8\textsuperscript{th} Xinyu Qiu}
\IEEEauthorblockA{\textit{Northeastern University} \\
Boston, USA \\
qiu.xiny@northeastern.edu}
}

\maketitle

\begin{abstract}
This paper presents a comprehensive performance evaluation of Large Language Models (LLMs) in solving programming challenges from Leetcode, a widely used platform for algorithm practice and technical interviews. We began by crawling the Leetcode website to collect a diverse set of problems encompassing various difficulty levels and topics. Using this dataset, we generated solutions with multiple LLMs, including GPT-4 and GPT-3.5-turbo (ChatGPT-turbo). The generated solutions were systematically evaluated for correctness and efficiency. We employed the pass@k metric to assess the success rates within a given number of attempts and analyzed the runtime performance of the solutions. Our results highlight the strengths and limitations of current LLMs\cite{10263390} in code generation and problem-solving tasks, providing insights into their potential applications and areas for improvement in automated programming assistance.

\end{abstract}


\section{Introduction}
Large Language Models (LLMs) like ChatGPT(OpenAI, 2023) have revolutionized artificial intelligence, demonstrating remarkable capabilities in text and image generation. In software development, specialized code-focused LLMs—such as CodeGen(Nijkamp etal., 2022), StarCoder(Li etal., 2023), WizardCoder(Luo etal., 2023), CodeT5(Wang etal., 2021), and Incoder(Fried etal., 2022)—assist developers by automating tasks like code generation, documentation, and unit testing. Additionally, LLMs have been integrated into Integrated Development Environments (IDEs) as code assistants\cite{10055297}, including GitHub Copilot,¹ Amazon CodeWhisperer,² and Tabnine.³ These tools aim to accelerate development by providing real-time code suggestions and automating routine coding tasks .
The integration of LLMs into software development offers significant benefits. Developers can save time, focus on higher-level design decisions, and potentially reduce time to market. LLMs help generate boilerplate code, suggest improvements, and assist with complex problem-solving, enhancing productivity and fostering innovation by leveraging the vast knowledge embedded within these models \cite{10735529}.
Despite their widespread adoption, research is increasingly focused on understanding the limitations and evaluating the performance of LLMs. Studies have highlighted security vulnerabilities in AI-generated code(Pearce etal., 2022; Sandoval etal., 2023; Perry ., 2023) and the prevalence of bugs(Jesse, 2023), emphasizing the need for thorough code review and testing. Other research explores how developers interact with LLMs\cite{tao2019fact} and integrate them into their workflows(Vaithilingam., 2022; Barke., 2023), examining the dynamics between human creativity and AI assistance\cite{tao2024nevlp}.
However, evaluating the runtime performance of LLM-generated code has received less attention. While correctness is crucial, the efficiency of code—how fast it runs and how optimally it uses resources—is a significant concern in software engineering\cite{li2024phishing}. Performance optimization is essential when resources are limited, scalability is needed, or energy consumption is a concern(Verdecchia., 2017; Acar etal., 2016). In areas like high-frequency trading, real-time data processing, or large-scale web applications, even minor execution time improvements can have substantial impacts. \cite{10754491}
To address this gap, our research evaluates the performance of code generated by LLMs on algorithmic challenges typical of programming contests and technical interviews. We conduct a comprehensive performance review using problems from Leetcode,⁴ a widely used platform offering a vast repository of algorithmic problems across various difficulty levels and topics.\cite{aghapour2024piqi}
Our key contributions are:
1. Performance Analysis of LLM-Generated Code: We analyze the performance of code generated by 18 LLMs on 204 Leetcode problems, investigating performance differences across models using a novel method for measuring and comparing runtime efficiency\cite{tao2023meta}.
2. Comparison with Human-Written Code: We compare the performance of LLM-generated code with human-written code, providing insights into the current capabilities of LLMs in producing efficient code and highlighting areas where they may lag behind human expertise.
3. Evaluation of Leetcode as a Dataset: We assess the usability of Leetcode as a public repository of algorithmic problems for research purposes, discussing its strengths and limitations to guide future research utilizing similar resources\cite{wang2018singlecaffe}.
Our methodology involves generating solutions using multiple LLMs, including GPT-4 and GPT-3.5-turbo, and systematically assessing their correctness and efficiency. We utilize metrics such as the pass@k metric, which evaluates the probability of a model providing a correct solution within \( k \) attempts, and measure the runtime performance of the generated code. \cite{wang2024low}
By analyzing these metrics, we aim to understand the strengths and limitations of current LLMs in algorithmic problem-solving contexts. Our findings offer insights into how LLMs can assist developers in tackling complex programming challenges and identify areas where further advancements are needed to enhance their capabilities in generating efficient, high-performance code\cite{wang2019dynamic}.

\section{Experiment}

This section outlines the experimental framework employed to evaluate the performance of Large Language Models (LLMs) in solving algorithmic problems from Leetcode. The experiment is structured into three primary phases: data collection, code generation, and solution evaluation.

\subsection{Data Collection}
To establish a comprehensive dataset for our evaluation, we crawled the Leetcode website and collected a total of 2,014 problems. These problems span various difficulty levels—Easy, Medium, and Hard—and encompass a wide range of topics including data structures, algorithms, mathematics, and system design.
During data collection, we focused on extracting the essential components necessary for code generation:

- Problem Statements: The detailed descriptions of each problem, including the objective and any specific requirements.

- Function Signatures: The provided code frameworks or templates, specifying input and output formats.

- Code Comments: Any comments included in the code templates that provide additional guidance or constraints.

We standardized the problem data by removing any extraneous information such as solution discussions, hints, or previously submitted solutions. This preprocessing ensured that the input to the LLMs was consistent and contained only the information that a typical developer would have when attempting to solve the problem independently.

\subsection{Code Generation}
The code generation phase involved utilizing two categories of LLMs to generate solutions for the collected Leetcode problems: OpenAI Models and GitHub Copilot Model. For each problem, we generated solutions using these models under varying levels of randomness and creativity, controlled by the temperature parameter in the models' settings. The temperature parameter influences the diversity of the output, with higher values producing more varied and creative responses. We utilized a Python framework to automate the code generation process\cite{wang2024load}. This framework automatically sent requests to the OpenAI API, providing the standardized problem input (problem statement, code comments, and code framework) as prompts. The solutions returned by the LLM were then parsed and formatted into the required Leetcode submission format in Python code.

\begin{itemize}
    \item \textbf{Temperature Settings:} We used five different temperature values: 0.2, 0.4, 0.6, 0.8, and 1.0.
    \item \textbf{Solution Generation:} At each temperature setting, we generated 10 distinct solutions per model for each problem.
\end{itemize}

\begin{itemize}
    \item \textbf{OpenAI Models:} We interfaced with the OpenAI API, providing the standardized problem input (problem statement, code comments, and code framework) as prompts. We set the temperature parameter accordingly and generated multiple solutions by invoking the model repeatedly.
    \item \textbf{GitHub Copilot:} We integrated Copilot into a compatible code editor (e.g., Visual Studio Code) and input the problem's code framework. Copilot's suggestions were captured for each temperature setting by configuring its randomness settings if available or by inducing variability through prompt modifications.
\end{itemize}

By generating multiple solutions across different temperatures, we aimed to observe the impact of the temperature parameter on the correctness and efficiency of the generated code. This process also allowed us to assess the models' ability to produce diverse solutions and their propensity to generate optimal or suboptimal code under varying conditions\cite{hu2024fewshotlearningadaptiveweight}.

\subsection{Solution Evaluation}
The evaluation phase involved assessing the correctness and performance of the generated solutions by submitting them to Leetcode's online judge system. The Leetcode platform provides an automated environment that compiles and executes submitted code against a predefined set of test cases.

For each submitted solution, we collected the following metrics
\begin{itemize}

\item \textbf{Number of Unit Tests Passed}: The total number of test cases successfully passed by the solution.
\item \textbf{Overall status}: indicating whether the solution met all the problem requirements.
\item \textbf{Runtime}: The execution time of the solution, measured by Leetcode's evaluation system.
\item \textbf{Memory Usage}: The amount of memory consumed during execution.

\end{itemize}
The evaluation process was conducted systematically:

\begin{enumerate}
    \item \textbf{Automated Submission:} Solutions were programmatically submitted to Leetcode using their API or through automated scripting to ensure consistency and efficiency.\cite{coignion2024performance}
    \item \textbf{Data Recording:} All evaluation results were recorded in a structured format for subsequent analysis. This included capturing the raw output from Leetcode and parsing relevant information.
\end{enumerate}

The collected data enabled us to analyze several aspects of the models' performance:

\begin{itemize}
    \item \textbf{Success Rate (Pass@\(\ k \) Metric):} The probability of a model generating a correct solution within \( k \) attempts, considering the multiple solutions generated per problem.
    \item \textbf{Error Analysis:} Identification of common errors or misconceptions exhibited by the models, such as off-by-one errors, incorrect loop conditions, or misuse of data structures.
    \item \textbf{Runtime Performance:} Assessment of the efficiency of the solutions, with a focus on execution time and resource utilization.
\end{itemize}

By evaluating both correctness and performance, we aimed to understand not only whether the models could solve the problems but also how efficiently they could do so. This dual focus is critical in algorithmic problem-solving contexts, where optimal solutions are often required to meet time and space constraints.

\subsection{Data Analysis}

In this section, we present the methods and metrics used to analyze the functional correctness and performance of the code generated by the Large Language Models (LLMs). Our analysis aims to assess not only whether the models can produce correct solutions but also how efficiently these solutions run compared to human-written code\cite{shen2024parameter}.

\subsubsection{D.1 Functional Correctness}

Functional correctness measures the extent to which the code generated by an LLM adheres to the specified problem requirements, effectively conforming to the ``program contract'' defined by the input prompt. To evaluate this aspect, we employed the pass@\(\ k \) metric, which calculates the probability that at least one of the \( k \) generated samples passes all the test cases for a given problem\cite{li2024deepreinforcementlearningbasedobstacle}.

We computed the pass@\(\ k \) metrics for \( k = 1 \) and \( k = 10 \), utilizing the unbiased estimator proposed by Chen \textit{et al.}~(2021). This estimator accounts for the likelihood of obtaining a correct solution among multiple attempts and is defined as:
\[
\text{pass@}k = \mathbb{E}\left[1 - \frac{\dbinom{n - c}{k}}{\dbinom{n}{k}}\right],
\]
where:
\begin{itemize}
    \item \( n \) is the total number of generated samples,
    \item \( c \) is the number of correct samples (i.e., samples that pass all test cases),
    \item \( \mathbb{E} \) denotes the expected value.
\end{itemize}
This formula provides an unbiased estimate of the pass@\(\ k \) metric by considering all possible combinations of correct and incorrect samples without replacement.

Following the methodology suggested by Chen \textit{et al.}~(2021), we calculated the pass@\(\ k \) for each temperature setting when evaluating an LLM's functional correctness. The temperature parameter influences the randomness and diversity of the generated solutions. By evaluating across different temperatures, we aimed to identify the optimal setting for each model. The best pass@\(\ k \) value observed across all temperatures was then considered the final pass@\(\ k \) metric for that LLM.

\subsubsection{D.2 Code Performance}

To assess the performance of the code generated by the LLMs, we considered three key metrics:
\begin{enumerate}
    \item \textbf{Memory Usage:}
    \begin{itemize}
        \item We recorded the memory consumption reported by Leetcode's evaluation system for each submitted solution. Memory usage is a critical factor in code performance, especially for problems with large input sizes or when operating under memory constraints.
    \end{itemize}
    \item \textbf{Runtime Performance:}
    \begin{itemize}
        \item We measured the execution time of the generated solutions using \texttt{pytest-benchmark}, a Python benchmarking tool. For each solution, we conducted multiple runs to obtain a reliable estimate of its runtime performance. The median runtime was computed to mitigate the impact of outliers and variability in execution times.
    \end{itemize}
    \item \textbf{Leetcode Runtime Percentile Rank:}
    \begin{itemize}
        \item Upon submission, Leetcode provides a percentile ranking that indicates how a solution's runtime compares to other users' submissions for the same problem. This rank is a value between 0 and 100, representing the percentage of submissions that the current solution outperforms. For example, a rank of 90 implies that the solution is faster than 90\% of all other submitted solutions. This metric allowed us to benchmark the LLM-generated code against human-written code at a global scale.
    \end{itemize}
\end{enumerate}
\section{Results}

As show in TABLE 1, which presents the performance of various AI models in pass-k metrics, likely representing different tasks or evaluation benchmarks. Below is an analysis of the data: $https://github.com/DHUer/LLM_evaluation_results$
\subsection{ Dataset analysis}
Our dataset analysis encompasses approximately 2,100 LeetCode problems, meticulously selected to provide a comprehensive evaluation of Large Language Models (LLMs) across a diverse range of algorithmic challenges. These problems are systematically categorized into three difficulty levels: Easy, Medium, and Hard, adhering to a distribution ratio of approximately 11:50:10, respectively.  Furthermore, all solutions generated by the LLMs were implemented in Python, a language renowned for its readability and widespread use in coding competitions and technical interviews. Additionally, each problem was approached using LLMs configured with five different temperature settings—0.2, 0.4, 0.6, 0.8, and 1.0. The temperature parameter controls the creativity and variability of the generated solutions, allowing us to examine how different levels of randomness impact the correctness and efficiency of the code produced. All the experiment code and dataset is published at: 

\subsection{LLMs solution compared with Humans}
To facilitate a robust comparison between LLM-generated solutions and human-written code,we selected the o1-mini model tested on LeetCode for this analysis. The results of this comparison are depicted in Figure 3. Utilizing LeetCode's runtime percentile rankings—which assume that the majority of historical submissions originate from human programmers—we assessed the execution speed of the LLM-generated solutions relative to human-written counterparts. Our findings reveal that the solutions produced by the selected LLM achieve a mean runtime percentile rank of 63\%, indicating that they are faster than 63\% of all previous submissions.

\subsection{Performance Overview}
\subsubsection*{Top Performers}
\begin{itemize}
    \item \textbf{Canonical Solutions} is the highest-performing model with near-perfect scores (97.94 and 98.04). This suggests it is tailored or highly optimized for the specific tasks.
    \item \textbf{GTP-4-omni}, \textbf{GPT-4}, and \textbf{GPT-4-turbo} follow but with significantly lower scores, indicating a strong performance but a noticeable gap compared to Canonical Solutions.
\end{itemize}

\subsubsection*{Mid-Tier Performers}
\begin{itemize}
    \item Models such as \textbf{Copilot}, \textbf{CodeLlama-13B-Instruct}, and \textbf{WizardCoder-Python-7B} show moderate performance (scores in the range of $\sim$4--19). This reflects some utility but highlights significant room for improvement compared to the top-tier models.
\end{itemize}

\subsubsection*{Lower Performers}
\begin{itemize}
    \item Models like \textbf{SantaCoder}, \textbf{InCoder-6B}, and \textbf{CodeT5-Large-NTP-PY} perform poorly with scores often below 5. These results suggest limited capability in handling the evaluated tasks effectively.
\end{itemize}
\begin{table}[h]
\centering
\caption{Pass@1 and Pass@10 Metrics for Various Models}
\label{tab:passk}
\begin{tabular}{lcc}
\hline
\textbf{Model}                       & \textbf{Pass@1 (\%)} & \textbf{Pass@10 (\%)} \\
\hline
Canonical Solutions                  & 97.94                & 98.04                 \\
GTP-4-omni                           & 43.36                & 61.95                 \\
GPT-4                                & 31.68                & 67.79                 \\
GPT-4-turbo                          & 31.25                & 50.00                 \\
Copilot                              & 8.09                 & 19.12                 \\
CodeLlama-13B-Instruct               & 4.17                 & 14.71                 \\
WizardCoder-Python-7B                & 4.17                 & 12.25                 \\
CodeLlama-13B-Python                 & 3.58                 & 14.71                 \\
CodeLlama-7B-Instruct                & 3.24                 & 12.75                 \\
StarCoder                            & 2.70                 & 10.29                 \\
CodeLlama-7B-Python                  & 2.60                 & 10.78                 \\
CodeLlama-7B                         & 2.11                 & 10.29                 \\
CodeGen2-7B-Instruct                 & 2.16                 & 10.78                 \\
CodeGen2-7B-Mono                     & 1.28                 & 7.35                  \\
CodeGen-6B-Mono                      & 1.08                 & 5.39                  \\
CodeGen-2B-Mono                      & 1.13                 & 6.37                  \\
Replit-Code-v1-3B                    & 0.98                 & 4.90                  \\
SantaCoder                           & 0.69                 & 4.90                  \\
InCoder-6B                           & 0.59                 & 3.92                  \\
InCoder-1B                           & 0.10                 & 0.98                  \\
CodeGen-350M-Mono                    & 0.39                 & 2.94                  \\
CodeT5-Large-NTP-PY                  & 0.25                 & 1.96                  \\
\hline
\end{tabular}
\end{table}

\begin{figure}
    \centering
    \includegraphics[width=1\linewidth]{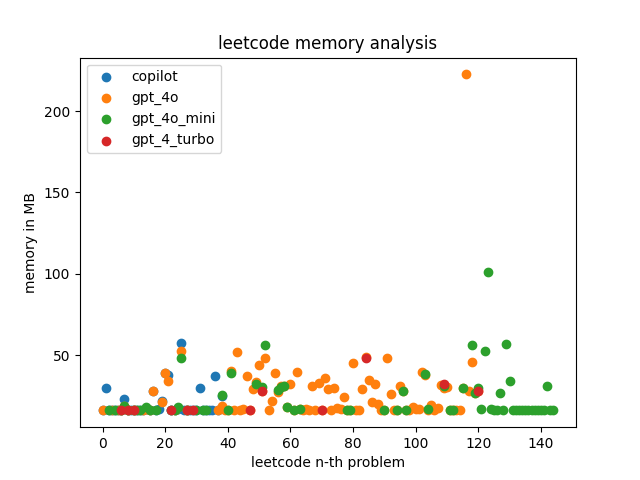}
    \caption{Leetcode memory analysis}
    \label{fig:enter-label}
\end{figure}

\begin{figure}
    \centering
    \includegraphics[width=1\linewidth]{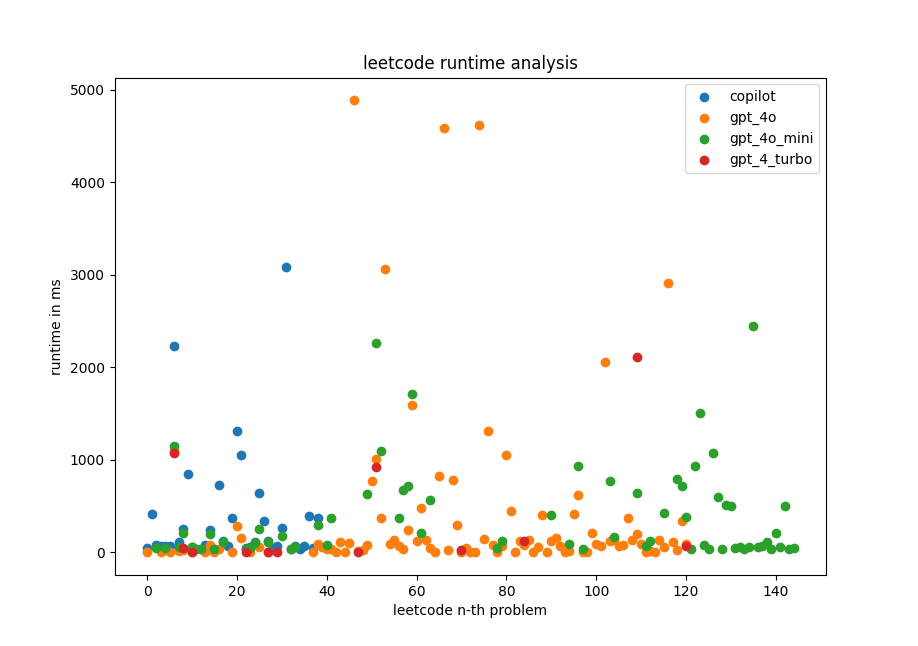}
    \caption{Leetcode runtime analysis}
    \label{fig:enter-label}
\end{figure}

\begin{figure}
    \centering
    \includegraphics[width=1\linewidth]{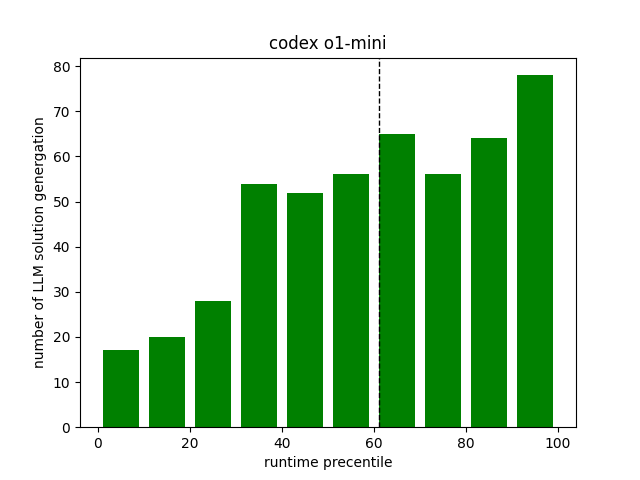}
    \caption{Distribution of the ranking for the runtime of the leetcode solution of o1-mini}
    \label{fig:enter-label}
\end{figure}
\bibliographystyle{plain}
\bibliography{citation}

\begin{thebibliography}{10}

\bibitem{aghapour2024piqi}
Ehsan Aghapour, Yixian Shen, Dolly Sapra, Andy Pimentel, and Anuj Pathania.
\newblock Piqi: Partially quantized dnn inference on hmpsocs.
\newblock In {\em Proceedings of the 29th ACM/IEEE International Symposium on Low Power Electronics and Design}, pages 1--6, 2024.

\bibitem{coignion2024performance}
Tristan Coignion, Cl{\'e}ment Quinton, and Romain Rouvoy.
\newblock A performance study of llm-generated code on leetcode.
\newblock In {\em Proceedings of the 28th International Conference on Evaluation and Assessment in Software Engineering}, pages 79--89, 2024.

\bibitem{10055297}
Shaoshuai Du, Kuangrong Hao, Haichao Zhang, Xue-song Tang, and Bing Wei.
\newblock Patch elastic deformation: An effective data augmentation method.
\newblock In {\em 2022 China Automation Congress (CAC)}, pages 2079--2084, 2022.

\bibitem{10735529}
Yiru Gong, Qimin Zhang, Huili Zheng, Zheyan Liu, and Shaohan Chen.
\newblock Graphical structural learning of rs-fmri data in heavy smokers.
\newblock In {\em 2024 4th International Conference on Computer Science and Blockchain (CCSB)}, pages 434--438, 2024.

\bibitem{hu2024fewshotlearningadaptiveweight}
Jiacheng Hu, Zhen Qi, Jianjun Wei, Jiajing Chen, Runyuan Bao, and Xinyu Qiu.
\newblock Few-shot learning with adaptive weight masking in conditional gans, 2024.

\bibitem{li2024phishing}
Daoming Li, Qiang Chen, and Lun Wang.
\newblock Phishing attacks: Detection and prevention techniques.
\newblock {\em Journal of Industrial Engineering and Applied Science}, 2(4):48--53, 2024.

\bibitem{li2024deepreinforcementlearningbasedobstacle}
Keqin Li, Jiajing Chen, Denzhi Yu, Tao Dajun, Xinyu Qiu, Lian Jieting, Sun Baiwei, Zhang Shengyuan, Zhenyu Wan, Ran Ji, Bo~Hong, and Fanghao Ni.
\newblock Deep reinforcement learning-based obstacle avoidance for robot movement in warehouse environments, 2024.

\bibitem{10754491}
Zheyan Liu, Qimin Zhang, Huili Zheng, Shaohan Chen, and Yiru Gong.
\newblock A comparative study of machine learning approaches for diabetes risk prediction: Insights from shap and feature importance.
\newblock In {\em 2024 5th International Conference on Machine Learning and Computer Application (ICMLCA)}, pages 35--38, 2024.

\bibitem{shen2024parameter}
Yixian Shen, Qi~Bi, Jia-Hong Huang, Hongyi Zhu, and Anuj Pathania.
\newblock Parameter-efficient fine-tuning via selective discrete cosine transform.
\newblock {\em arXiv preprint arXiv:2410.09103}, 2024.

\bibitem{10263390}
Yiyi Tao.
\newblock Meta learning enabled adversarial defense.
\newblock In {\em 2023 IEEE International Conference on Sensors, Electronics and Computer Engineering (ICSECE)}, pages 1326--1330, 2023.

\bibitem{tao2023meta}
Yiyi Tao.
\newblock Meta learning enabled adversarial defense.
\newblock In {\em 2023 IEEE International Conference on Sensors, Electronics and Computer Engineering (ICSECE)}, pages 1326--1330. IEEE, 2023.

\bibitem{tao2019fact}
Yiyi Tao, Yiling Jia, Nan Wang, and Hongning Wang.
\newblock The fact: Taming latent factor models for explainability with factorization trees.
\newblock In {\em Proceedings of the 42nd international ACM SIGIR conference on research and development in information retrieval}, pages 295--304, 2019.

\bibitem{tao2024nevlp}
Yiyi Tao, Zhuoyue Wang, Hang Zhang, and Lun Wang.
\newblock Nevlp: Noise-robust framework for efficient vision-language pre-training.
\newblock {\em arXiv preprint arXiv:2409.09582}, 2024.

\bibitem{wang2018singlecaffe}
Chenxu Wang, Yixian Shen, Jia Jia, Yutong Lu, Zhiguang Chen, and Bo~Wang.
\newblock Singlecaffe: an efficient framework for deep learning on a single node.
\newblock {\em IEEE Access}, 6:69660--69671, 2018.

\bibitem{wang2024low}
Lun Wang.
\newblock Low-latency, high-throughput load balancing algorithms.
\newblock {\em Journal of Computer Technology and Applied Mathematics}, 1(2):1--9, 2024.

\bibitem{wang2024load}
Lun Wang, Wei Fang, and Yudi Du.
\newblock Load balancing strategies in heterogeneous environments.
\newblock {\em Journal of Computer Technology and Applied Mathematics}, 1(2):10--18, 2024.

\bibitem{wang2019dynamic}
Lun Wang, Wentao Xiao, and Shan Ye.
\newblock Dynamic multi-label learning with multiple new labels.
\newblock In {\em Image and Graphics: 10th International Conference, ICIG 2019, Beijing, China, August 23--25, 2019, Proceedings, Part III 10}, pages 421--431. Springer, 2019.

\end{thebibliography}

\end{document}